\documentclass[12pt]{dcds2}

\usepackage{graphicx}
\usepackage{amssymb}

\def\ppages{1--10}

\newcommand{\be}{\begin{equation}}
\newcommand{\ee}{\end{equation}}
\newcommand{\bea}{\begin{eqnarray}}
\newcommand{\eea}{\end{eqnarray}}

\theoremstyle{plain}

\theoremstyle{definition}

\title[]
{Nonlinear Dynamics In Quantum Physics -- Quantum Chaos and Quantum Instantons}

\email{hkroger@phy.ulaval.ca}

\author[H.~Kr\"{o}ger]{}

\begin{document}
\maketitle

\centerline{\scshape Helmut Kr\"{o}ger}

\medskip

{\footnotesize \centerline{} 
\centerline{D\'{e}partement de Physique}
\centerline{Universit\'{e} Laval}
\centerline{Qu\'{e}bec, Qu\'{e}bec G1K 7P4, Canada}
 }

\begin{quote}{\normalfont\fontsize{8}{10}\selectfont
{\bfseries Abstract.}
We discuss the recently proposed quantum action - its interpretation, its motivation, its mathematical properties and its use in physics: 
quantum mechanical tunneling, quantum instantons and quantum chaos.
\par}
\end{quote}

\section{Introduction}
\label{sec:Intro} 

Modern physics returns to some of its origins dating back to the first part of the last century. Examples are entanglement, according to Schr\"odinger the most peculiar property occuring in quantum mechanics, or the condensation of very cold atoms predicted by Einstein and Bose (Bose-Einstein condensate).
Another example is nonlinear dynamics and chaos, dating back to the work of Poincar\'e and others and its modern descendents - quantum chaos.
Entanglement plays a fundamental role in quantum optics, quantum computing, quantum cryptography, and quantum teletransportation \cite{Macchiavello,Bouwmeester}. The Bose-Einstein condensate is a new state of matter. Classical chaos plays an important role in macroscopic physics, chemistry and biology. Quantum chaos may become quite important for example in semiconductor devices, when due to decreasing size (from micron to Angstrom) atomic length scales are reached and the laws of quantum mechanics apply. Like chaotic motion in a fluid, quantum chaos may cause a similar phenomenon for the flux of the electric current in a semiconductor.

What has the quantum action - the subject of this article - to do with the above issues? The quantum action is a new concept concerning the foundations of quantum mechanics. Since de Broglie we know that the quantum world has apparently a dual face - a particle picture and a wave picture.   
The quantum action can be considered as a link between classical physics and quantum physics. This may be interpreted as if the quantum world has a third face - a classical picture with a new interaction. 
Consequently, the quantum action may be expected to be useful to better understand the relation between and the transition from classical physics to quantum physics. In particular, there are some very fruitful concepts in quantum physics, which are derived from classical physics: One example is the instanton solution in a quantum field theory. Another example is quantum chaos.

On the other hand there are phenomena occuring in the quantum world which do not have a counter part in the classical world. A well known example is tunneling:
A double well potential with a barrier in the middle prohibits in classical physics a transition from one minimum to the other if the energy is too small to overcome the barrier. However, in quantum physics a transition does occur with a certain probability. It is noteworthy that tunneling and instantons are closely related \cite{Coleman}.
There are other systems, where transitions are classically forbidden, due to some conserved quantity (other than energy), but quantum mechanical transitions are possible. This is known as dynamical tunneling \cite{Anderson}. It has recently been realized experimentally \cite{Raizen,Hensinger}. It turns out that quantum chaos seems to facilitate dynamical tunneling. 
We suggest that the quantum action \cite{Jirari01a,Jirari01b,Jirari01c,Caron01,Kroger02} 
is an appropriate tool to investigate 
phenomena on the borderline between classical and quantum physics and to analyze   quantum tunneling, quantum instantons and quantum chaos.

In the following we discuss the concept of the quantum action. In particular, we discuss analytical properties of the quantum action in the limit when the time of a quantum mechanical transition goes to infinity.
As an example of the use of the quantum action we discuss 
quantum instantons from a double well potential and quantum chaos from the 2-D anharmonic oscillator.
Finally, we give an outlook on further use of the quantum action.

\section{The problem of quantum chaos and quantum instantons}
\label{sec:ProblQuantChaos}

Chaos is quite well understood in classical physics. 
Its origin in terms of nonlinear dynamics has been investigated thoroughly 
in theoretical physics and mathematics. 
If one considers nature at the level of atoms, where the laws of quantum mechanics (Q.M.) hold, chaotic phenomena are by far less well understood than 
in classical (macroscopic) physics \cite{Blumel,Gutzwiller,Haake,Nakamura,Stockmann}. 
Quantum chaos is usually analyzed in terms of the level density of its energy spectrum. This is based on a conjecure by Bohigas et al. \cite{Bohigas}
which states that the level density distribution (Poissonian vs. Wignerian) 
corresponds to an integrable (non-chaotic) vs. chaotic quantum system. 
In classically chaotic systems, phase space portraits, Poincar\'e sectioons,  
Lyapunov exponents and the KAM theorem are very useful to analyze the chaotic behavior. However, those notions have no direct counterpart in quantum systems. 
The problem is due to the fact that in quantum mechanics one can not specify with certainty position and momentum at the same time (Heisenberg's uncertainty relation). Hence a sharp point in phase space has no meaning. 

This is also the problem with the definition of quantum instantons.
Recall that in classical physics the instanton is a solution in imaginary time where a particle moves in a potential with degenerate minima, such that it starts at one minimum
(sharp position and velocity zero) and goes over to another minimum (sharp position and velocity zero). Again boundary conditions are sharp points in phase space with no direct analogon in Q.M.  

Here we take the following approach: Via the quantum action one can parametrize quantum mechanical transition amplitudes (which contain all information about the physics of the system). On the other hand, the quantum action has the same mathematical form as the classical action, a kinetic term 
(proportional to velocity squared) and a local potential term. However, the action parameters, like mass and potential are different from the classical one, in general. Starting from such quantum action then allows to construct a phase space portrait and hence reintroduce the tools of classical chaos theory to the quantum world.

Let us come back to the problem of a point in phase space used as initial condition. 
Recall in classical mechanics that equations of motions
are fixed by specifying the dynamics (usually in terms of a differential equation) and by specifying initial conditions. Usually one specifies position and velocity at some initial time. 
A consequence of this type of boundary conditions is that certain trajectories are forbidden. For example those where the energy is too small to overcome a potential barrier. Moreover, this initial condition has no direct analogon in Q.M.
However, there is a simple way out: Consider classical mechanics with 2-point boundary conditions. I.e., one specifies positions at some initial and final time. 
Then there are no classically forbidden trajectories (sometimes even infinitely many trajectories are possible).
In quantum physics, there is a well defined transition amplitude. Below we define the quantum action by use of two-point boundary conditions.

\section{Postulate of quantum action}
\label{sec:Postulate}  

The quantum action \cite{Jirari01a} is defined by the following requirements.
For a given classical action 
\begin{equation}
S[x] = \int dt \frac{m}{2} \dot{x}^{2} - V(x) ~ ,
\end{equation}
there is a quantum action
\begin{equation}
\tilde{S}[x] = \int dt \frac{\tilde{m}}{2} \dot{x}^{2} - \tilde{V}(x) ~ ,
\end{equation}
which parametrizes the Q.M. transition amplitude 
\begin{eqnarray*}
&& G(x_{f},t_{f}; x_{i},t_{i}) = \tilde{Z} 
\exp \left[ \frac{i}{\hbar} 
\left. \tilde{\Sigma} \right|_{x_{i},t_{i}}^{x_{f},t_{f}} \right] ~ , 
\nonumber \\
&& \left. \tilde{\Sigma} \right|_{x_{i},t_{i}}^{x_{f},t_{f}} 
= \left. \tilde{S}[\tilde{x}_{cl}] \right|_{x_{i},t_{i}}^{x_{f},t_{f}}  
= \left. \int_{t_{i}}^{t_{f}} dt ~ \frac{\tilde{m}}{2} \dot{\tilde{x}}_{cl}^{2} - \tilde{V}(\tilde{x}_{cl}) \right|_{x_{i}}^{x_{f}}  ~ .
\end{eqnarray*}
where $\tilde{x}_{cl}$ denotes the classical path corresponding to the action $\tilde{S}$. We require 2-point boundary conditions,
\begin{eqnarray*}
&& \tilde{x}_{cl}(t=t_{i}) = x_{i}
\nonumber \\
&& \tilde{x}_{cl}(t=t_{f}) = x_{f} ~ .
\end{eqnarray*}
$\tilde{Z}$ stands for a dimensionful normalisation factor. 
Note: The parameters of the quantum action (mass, potential) are independent of the boundary points $x_{f}$, $x_{i}$, but depend on the transition time  
$T=t_{f}-t_{i}$. The same q-action parametrizes all transition amplitudes for a given transition time. 

The postulate leads to the question: Does such quantum action exist?
So far we have no general proof. However, the answer is affirmative in the following cases:
\newline (a) Harmonic oscillator. In this case the classical action satisfies the definition of the quantum action \cite{Schulman,Jirari01a}, hence both coincide.
\newline (b) Arbitrary local potential, limit when transition time $T \to 0$: 
Then the quantum action exists as it coincides with classical action. 
\newline (c) Arbitrary potential, imaginary time, limit when transition time  
$T \to \infty$ (equivalent to temperature going to zero): 
Then the quantum actions exists, being different from the classical action, in general (a proof is given in Ref.\cite{Kroger02}).

\section{Analytical properties of the quantum action} 
\label{sec:Analytical}

In this limit of large imaginary time, the quantum action has a number of remarkable properties: \\
(i) The WKB approximation for the ground state wave function becomes exact, after replacing the classical action by the quantum action. \\
(ii) There is an analytic expression for the ground state wave function in terms of the quantum action. \\
(iii) The ground state energy coincides with the minimum of the quantum potential. \\
(iv) The ground state wave function has a maximum exactly at the same position, where the quantum potential has a minimum. \\
(v) The quantum action allows also to reproduce energies and wave functions of excited states. Example: The spectrum of the hydrogen atom, considering the lowest energy states for given angular momentum. \\
Let us detail some of those results.   

For physical and mathematical reasons, it is interesting to go from real time over to imaganiary time. A mathematical reason is that the path integral for the Q.M. transition amplitude then becomes well defined (Wiener integral). A physical reason is that the description of finite temperature physics requires the use of imaginary time. Thus we make the following transition
\begin{equation*}
t \to -it ~ .
\end{equation*}
Then the action becomes the so-called Euclidean action 
\begin{equation}
S_{E} = \int_{0}^{T} dt \frac{m}{2} \dot{x}^{2} + V(x) ~ ,
\end{equation}
the transition amplitude becomes the Euclidean transition amplitude
\begin{eqnarray*}
G_{E}(x_{f},T;x_{i},0) &=& \langle x_{f} | \exp[ - H T/\hbar ] | x_{i} \rangle
\nonumber \\
&=& \left. \int [dx] \exp \left[ - S_{E}[x]/\hbar \right] \right|_{x_{i},0}^{x_{f},T} ~ ,
\end{eqnarray*}
and the quantum action becomes the Euclidean quantum action 
\begin{equation}
G_{E}(x_{f},T; x_{i},0) = \tilde{Z}_{E} 
\exp \left[ -\frac{1}{\hbar} 
\left. \tilde{\Sigma}_{E} \right|_{x_{i},0}^{x_{f},T} \right] ~ . 
\end{equation}

Let us recall the Feynman-Kac formula, which makes a statement about the time 
evolution of the quantum system in the limit of large imaginary time.
In this limit the time evolution of a quantum system is determined by 
the behavior of $\exp[- H T/\hbar]$ when $T$ goes to $\infty$. In particular, one has 
\begin{equation*}
e^{- H T/\hbar } 
\longrightarrow_{T \to \infty} 
|\psi_{gr} \rangle e^{- E_{gr} T/\hbar} \langle \psi_{gr} | ~ ,
\end{equation*}
where $E_{gr}$ and $\psi_{gr}$ are the ground state energy and wave function, respectively.

Now let us consider the case of a potential $V(x)$ having a unique minimum and obeying $V(x) \to \infty$ when $x \to \pm \infty$. We consider the Euclidean action in the limit of large time. 
The trajectory minimizes the action. Because the kinitic term and the potential term (by assumption) are positive, this means that the trajectory minimizes fluctuations and stays as close as possible to the bottom of the potential valley. This implies for $T \to \infty$,
\begin{equation*}
\tilde{V} \to \tilde{V}_{min} ~ , ~~~ 
\tilde{T}_{kin} \to 0 ~ , ~~~
\epsilon = - \tilde{T}_{kin} + \tilde{V} \to \tilde{V}_{min} ~ ,
\end{equation*}
and
\begin{equation*}
\tilde{\Sigma} \equiv \tilde{S}[\tilde{x}_{cl}]|_{x_{i},0}^{x_{f},T} 
= \int_{0}^{T} dt ~ \tilde{T}_{kin} + \tilde{V} 
= \tilde{v}_{0} T + \left( 
\int_{x_{i}}^{0} + \int_{0}^{x_{f}} dx ~  
_{\pm} \sqrt{2 \tilde{m}(\tilde{V}(x) -\tilde{V}_{min}) } 
\right) ~ ,
\end{equation*}
where the sign depends on initial and final data. 
The transition amplitude then becomes
\begin{eqnarray*}
&&G(x_{f},T;x_{i},0) = \left. \tilde{Z} \exp[- \tilde{\Sigma}/\hbar] \right|_{x_{i},0}^{x_{f},T} 
\stackrel{T \to \infty}{\longrightarrow} 
\tilde{Z}_{0} ~ \exp[ -\tilde{v}_{0} T/\hbar ]
\nonumber \\
&&\times
\exp[ - \int_{0}^{x_{fi}} dx ~ 
\sqrt{2 \tilde{m}( \tilde{V}(x) -\tilde{V}_{min} ) }/\hbar  ]
\exp[ - \int_{x_{in}}^{0} dx ~ 
\sqrt{2 \tilde{m}( \tilde{V}(x) - \tilde{V}_{min} ) }/\hbar ] ~ . 
\end{eqnarray*}
By comparison with the Feynman-Kac formula one obtains
the following analytic expressions for the ground state energy and wave function, expressed in terms of the quantum action,
\begin{equation}
\label{eq:QPotExtr}
E_{gr} = \tilde{V}_{min} ~ , ~~~ 
\psi_{gr}(x) = \frac{1}{N} ~ e^{ - \int_{0}^{|x|} dx' ~ 
\sqrt{2 \tilde{m}( \tilde{V}(x') - \tilde{V}_{min} ) }/\hbar } ~ .
\end{equation}
Combining this with the Schr\"odinger equation leads to the following transformation law
\begin{equation}
\label{eq:TransLaw}
2 m(V(x) - E_{gr})  
=2 \tilde{m}(\tilde{V}(x) - \tilde{V}_{min}) 
- \frac{\hbar}{2} \frac{ \frac{d}{dx} 2 \tilde{m} (\tilde{V}(x) - \tilde{V}_{min})}
{ \sqrt{2 \tilde{m}( \tilde{V}(x) - \tilde{V}_{min} ) } } ~ \mbox{sgn}(x) ~ .
\end{equation}
Although those results have been obtained in imaginary time,
they hold also in real time (ground state energy $E_{gr}$ and 
wave function $\psi_{gr}(x)$ are the same as in real time).

\begin{figure}[thb]
\vspace{9pt}
\begin{center}
\includegraphics[scale=0.4,angle=0]{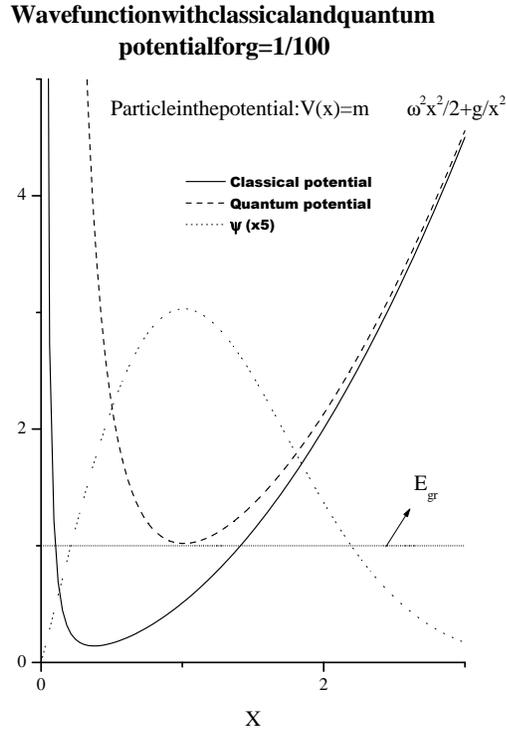}
\end{center}
\caption{Inverse square potential. Classical potential, quantum potential and groundstate wave function.}
\label{Fig1}
\end{figure}

Eq.(\ref{eq:QPotExtr}) shows that 
the minimum of the quantum potential coincides with the ground state energy.
Second, the position of the minimum of the quantum potential coincides with the position of the maximum of the ground state wave function. This is shown in Fig.[\ref{Fig1}]. Hence the quantum potential gives a much better picture of the behavior of the quantum system than the classical potential, which does not share those properties.

The celebrated semi-classical WKB formula [Wentzel, Kramers and Brillouin (1926)] gives approximate solutions to wave function and tunneling amplitude. 
As an example let us consider in 1-D a system with parity symmetric "confinement" potential (potential goes to $+ \infty$ when $|x| \to \infty$). 
Assume that the system is in the ground state. 
The WKB formula gives an approximate expression for wave function at energy $E=E_{gr}$
\begin{equation*}
\psi_{WKB}(x) = \frac{A}{(2m[V(x) - E_{gr}])^{1/4}} ~
\exp\left[ - \frac{1}{\hbar} \int_{x_{0}}^{x} dx' \sqrt{2m[V(x') - E_{gr}]} \right] 
~ .
\end{equation*}
Comparing with the results from sect.\ref{sec:Analytical},
we see that the WKB formula becomes exact, when replacing 
parameters of classical action by those of the quantum action
\begin{equation*}
m \to \tilde{m} ~ , ~~~ 
V(x) \to \tilde{V}(x) ~ , 
\end{equation*}
and replacing the rational term in front by a constant (wave function normalisation).

\section{Excited states of hydrogen atom} 
\label{sec:Hydrogen}

Can the quantum action also give analytical results for excited states?
The answer is yes, if we consider excited states being lowest energy states of a conserved quantum number.
As an example let us consider the radial motion of the hydrogen atom in a sector of fixed angular momentum $l>0$. 
The potential has a centrifugal plus a Coulomb term. 
Let us consider angular motion to be quantized.
We keep the angular momentum quantum number $l$ fixed. 
In the Feynman-Kac limit the transition amplitude is projected onto the state of lowest energy compatible with quantum number $l$. 
The states of the hydrogen atom are characterized 
by the quantum numbers $n$ (principal quantum number) $l$ (angular momentum) where $n=l+1$. There is also the magnetic quantum number $m$.
The radial potential is given by
\begin{equation*}
V_{l}(r) = \frac{\hbar^{2}l(l+1)}{2m r^{2}} - \frac{e^{2}}{r} ~ .
\end{equation*}
The exact energy of the excited states is given by
\begin{equation*}
E_{l} = - \frac{E_{I}}{n^{2}}= - \frac{E_{I}}{(l+1)^{2}} ~ , ~~~ 
E_{I} = \frac{m e^{4}}{2 \hbar^{2}} \approx 13.6 eV ~ \mbox{(ionisation energy)} ~ .
\end{equation*} 
The corresponding wave function is given by
\begin{equation*}
\phi_{l}(r) = \frac{1}{N_{l}} ~ \left(\frac{r}{a_{0}}\right)^{l} ~ 
\exp\left[ -\frac{r}{(l+1) a_{0}} \right] ~ , ~~~
a_{0} = \frac{\hbar^{2}}{m e^{2}} ~ \mbox{(Bohr radius)} ~ . 
\end{equation*}
For the quantum action we make an ansatz 
\begin{equation*}
\tilde{m}=m ~ , ~~~
\tilde{V}_{l}(r) = \mu/r^{2} -\nu/r ~ .
\end{equation*}
A transformation law [similar to Eq.(\ref{eq:TransLaw})] determines the parameters of the q-action,
\begin{equation*}
\mu = \frac{\hbar^{2}}{2m} l^{2} ~ , ~~~
\nu = e^{2} \frac{l}{l+1} ~ ,
\end{equation*}
and the minimum of the quantum potential gives exactly the excitation energies
\begin{equation*}
E_{l} = \tilde{V}^{min}_{l} = - \frac{m e^{4}}{2 \hbar^{2}} \frac{1}{(l+1)^{2}} = - \frac{E_{I}}{(l+1)^{2}} ~ .
\end{equation*}
Again the wave function can be expressed in terms of the quantum action in a way similar to Eq.(\ref{eq:QPotExtr}), and reproduces the exact wave function $\phi_{l}$.
Moreover, we observe that the excited state wave function $\phi_{l}(r)$ has its maximum where the quantum potential $\tilde{V}_{l}$ has its minimum.
The quantum action has the same structure as the classical action. 
Both, the centrifugal and the Coulomb term get tuned.

\section{Quantum instantons}
\label{sec:QuantInst}

\begin{figure}[thb]
\vspace{9pt}
\begin{center}
\includegraphics[scale=0.4,angle=0]{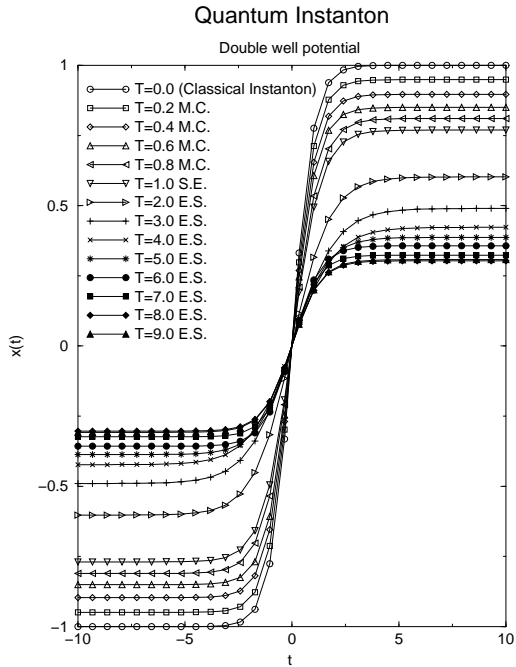}
\end{center}
\caption{Instantons in double well potential.}
\label{Fig2}
\end{figure}

We consider quantum mechanics in 1-D. A particle of mass 
$m$ moves in a quartic potential of double well shape given by 
\begin{equation*}
V(x) = \frac{1}{2} - x^{2} + \frac{1}{2} x^{4} ~ ,
\end{equation*}
and $m = 1$, $\hbar = 1$.
The potential minima lie at $a=\pm 1$. The potential barrier has the height $B=1/2$. The double well potential has a classical instanton solution. It is obtained by 
solving the Euler-Lagrange equations of motion of the Euclidean classical 
action, with the initial conditions $x(t=-\infty) = -a$, $\dot{x}(t=-\infty)=0$,
\begin{equation*}
\label{ClassInstanton}
x^{cl}_{inst}(t) = a ~\mbox{tanh}[ \sqrt{2/m}~ A ~a ~t] ,
\end{equation*}
where $A=1/\sqrt{2}$. The classical instanton goes from
$x(t=-\infty) = -a$ 
to $x(t=+\infty) = +a$ (see Fig.[\ref{Fig2}]).

For the quantum action we make the following ansatz,
\begin{equation*}
\label{QuantActionDoublWell}
\tilde{S} = \int dt \frac{1}{2} \tilde{m} \dot{x}^{2} - \sum_{k=0}^{4} \tilde{v}_{k} x^{k}.
\end{equation*}

For example, at (imaginary) transition time $T=0.5$ we find the following parameters for the quantum action
$\tilde{m}=0.9961(2)$,  
$\tilde{v}_{0}=1.5710(17)$, 
$\tilde{v}_{1}=0.000(2)$, 
$\tilde{v}_{2}=-0.745(6)$, 
$\tilde{v}_{3}=0.000(2)$, 
$\tilde{v}_{4}=0.493(3)$. 
By adding a constant, the quantum potential can be expressed as
a positive double well with minima located at $\pm \tilde{a}$ with 
$\tilde{a}=0.869(6)$. It has a barrier height $\tilde{B}=0.281$.
It displays "degenerate vacua". 
Compared with the classical potential it is softer, i.e. its potential minima are closer to the origin and its barrier is also lower.
The quantum potential also has an instanton solution,
corresponding to $T=0.5$, given by 
\begin{equation*}
x_{inst}^{T=0.5}(t) = \tilde{a} ~\mbox{tanh}[\sqrt{2/\tilde{m}} ~\tilde{A} ~\tilde{a} ~t] 
\approx 0.869 ~ \mbox{tanh} [0.865 ~ t ] .
\end{equation*}
Similarly, we find an instanton solution for any larger value of $T$. The 
quantum instanton is obtained in the asymptotic limit $T \to \infty$.
The evolution of the instantons under variation of $T$, i.e. under variation of the temperature, is depicted in Fig.[\ref{Fig2}]. It shows the transition from the classical instanton (at infinite temperature) to the quantum instanton
(at zero temperature). One observes that the quantum instanton is much softer than the classical instanton.

\section{Quantum chaos}
\label{sec:QuantChaos}

The hydrogen atom in the presence of strong magnetic fields has been explored experimentally and theoretically \cite{Bohigas,Friedrich, Wintgen}.
It displays quantum chaos via disorder in the spectrum. In the regime of high lying excited states, where the system becomes semi-classical, Gutzwiller's trace formula has been applied successfully by Wintgen \cite{Wintgen} to extract periodic orbit information. Here we suggest to explore quantum chaos 
via the phase space generated by the quantum action. As a prototype system, we consider a 2-dim non-integrable Hamiltonian, with a classically chaotic counter part. 

Our definition of quantum chaos is based on the concept of 'some' phase space related to quantum mechanics. The novel idea is to introduce this phase space via the quantum action. Second, the quantum action depends on the transition time $T$, however, for large $T$ it converges  
asymptotically. Moreover, in this regime the existence of the quantum action has been established rigorously. The regime of large $T$ makes physical sense, because the proper definition of Lyapunov exponents, one of the characteristics of chaotic dynamics, involves the large time limit. 

As an example, let us consider the following Hamiltonian
\begin{eqnarray*}
&& S = \int_{0}^{T} dt ~ \frac{1}{2} m (\dot{x}^{2} + \dot{y}^{2}) 
+ V(x,y) 
\nonumber \\
&& V(x,y) = v_{2}(x^{2} + y^{2}) + v_{22} x^{2}y^{2} 
\nonumber \\
&& m = 1 ~ , ~~~ 
v_{2} = 0.5 ~ , ~~~
v_{22} = 0.05 ~ .
\end{eqnarray*}
The parameter $v_{22}$ controls the deviation from integrability.
At $v_{22}=0$ the system is integrable. The system corresponds to two uncoupled oscillators, the corresponding quantum action then coincides with the classical action. One expects that tuning the parameter 
$v_{22}$, while keeping the other parameters fixed, the system will become more chaotic when increasing $v_{22}$. It is interesting to ask: Does the quantum system also become more chaotic? 

For the quantum action, we have made the following ansatz: It reflects the time-reversal symmetry, parity conservation and symmetry under exchange 
$x \leftrightarrow y$:
\begin{eqnarray*}
\tilde{S} &=& \int_{0}^{T} dt ~ 
\frac{1}{2} \tilde{m} (\dot{x}^{2} + \dot{y}^{2}) 
+ \tilde{V}(x,y), ~~~
\nonumber \\
\tilde{V} &=& \tilde{v}_{0} 
+ \tilde{v}_{2} (x^{2} + y^{2}) 
+ \tilde{v}_{22} x^{2}y^{2} 
+ \tilde{v}_{4} (x^{4} + y^{4}) ~ .
\end{eqnarray*}

We also have included in the quantum action terms like 
$\dot{x} \dot{y}$,
$xy$,
$xy^{3}+x^{3}y$,
$x^{2}y^{4} + x^{4}y^{2}$,
$x^{4}y^{4}$. 
Numerically, those coefficients were found to be small (compared to machine precision or zero within error bars).

\begin{figure}[thb]
\vspace{9pt}
\begin{center}
\includegraphics[scale=0.55,angle=0]{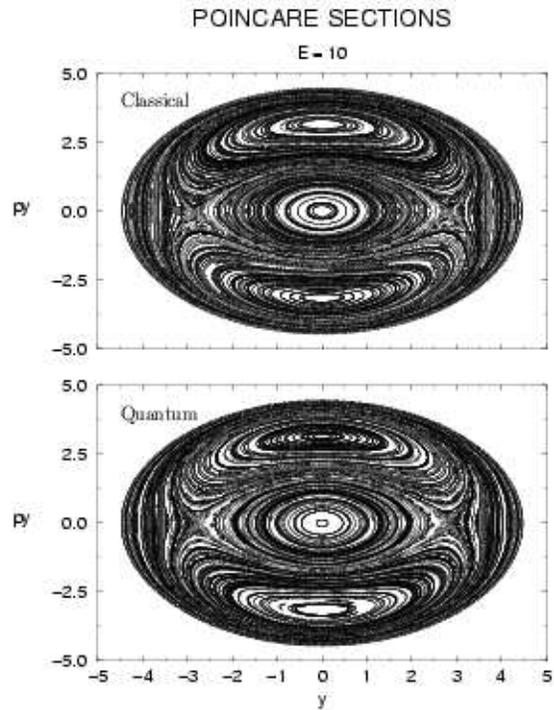}
\end{center}
\caption{2-D anharmonic oscillator. Poincar\'e section of classical action vs. quantum action.}
\label{Fig3}
\end{figure}

An example of Poincar\'e sections for the classical system and the quantum systen are shown in Fig.[\ref{Fig3}]. The quantum Poincar\'e section corresponds to transition time $T=4.5$. The energy has been chosen to be $E=10$. 
One can observe some differences between the classical and the quantum system.
Further numerical studies \cite{Caron01} have shown the following behavior:
For small $v_{22}$, Poincar\'e sections of classical and quantum physics are quite similar. With increase of energy, both display mixed dynamics and become more chaotic. Also, with increase of energy differences between classical and quantum phasse space become more pronounced. Islands of regular behavior differ in shape and position. Quantitatively, one observes that the value of 
$\tilde{v}_{22}$ (quantum action) is smaller than the corresponding value 
$v_{22}$ of the classsical action. Because this parametyers measures the deviation from integrability, this hints to the possibility that the quantum system is "less" chaotic than the classical system. A more detailed analysis is called for.

\section{Outlook: Further use of quantum action.}
\label{sec:Outlook}

In cosmology, the inflationary scenario describes the early universe. 
Inflation involves potentials with several minima and instanton solutions. 
The instanton starts out as a quantum instanton and eventually turns into a classical instanton. This has effects on the subsequent formation of galaxies.
The quantum action allows to search for new minima in the quantum potential, which may be absent in the classical potential. 

In condensed matter physics, 
superconducting quantum interference devices (SQUID)
have been used by Friedman et al. to demonstrate experimentally the phenomenon of quantum superposition in macroscopic states. This involves Josephson junctions. The SQUID potential has a double-well structure. 
The quantum action allows to construct a quantum potential and analyze 
quantum superposition in terms of such quantum potential.

In atomic physics, Steck et al. and Hensinger et al. 
have demonstrated experimentally the phenomenon of dynamical tunneling (where the classical transition is forbidden due to some conserved quantity different from energy). It has been realized by arrays of cold atoms. It has been observed that the presence of quantum chaos enhances the dynamical tunneling transition.
It would be interesting to reexamine dynamical tunneling using the phase space portrait constructed from the quantum action.

In chemistry, the process of chemical binding of macromolecules often involves potentials of a double well structure. 
The quantum action can be used to study the formation of such macromolecules. 
\\

\noindent {\bf Acknowledgements} \\ 
H.K. is grateful for support by NSERC Canada.

\end{document}